\documentclass[twocolumn]{article}
\usepackage{times}
\usepackage{graphicx}
\usepackage{amssymb}
\usepackage{amsmath}
\usepackage{amsfonts}
\usepackage[english]{babel}
\usepackage{booktabs}
 \usepackage{harvard}
 \citationmode{abbr}

\title{The Ising decoder: reading out the activity of large neural ensembles}
\author{Michael T Schaub and Simon R Schultz\\
{\it Department of Bioengineering, Imperial College London,} \\
{\it South Kensington, London SW7 2AZ, UK.} \\
{\ttfamily s.schultz@imperial.ac.uk}
}

\begin{document}

\maketitle

\begin{abstract}
The Ising Model has recently received much attention for the statistical description of neural spike train data. In this paper, we propose and demonstrate its use for building decoders capable of predicting, on a millisecond timescale, the stimulus represented by a pattern of neural activity. After fitting to a training dataset, the Ising decoder can be applied ``online" for instantaneous decoding of test data. While such models can be fit exactly using Boltzmann learning, this approach rapidly becomes computationally intractable as neural ensemble size increases. We show that several approaches, including the Thouless-Anderson-Palmer (TAP) mean field approach from statistical physics, and the recently developed Minimum Probability Flow Learning (MPFL) algorithm, can be used for rapid inference of model parameters in large-scale neural ensembles. Use of the Ising model for decoding, unlike other problems such as functional connectivity estimation, requires estimation of the partition function. As this involves summation over all possible responses, this step can be limiting. Mean field approaches avoid this problem by providing an analytical expression for the partition function. We demonstrate these decoding techniques by applying them to simulated neural ensemble responses from a mouse visual cortex model, finding an improvement in decoder performance for a model with heterogeneous as opposed to homogeneous neural tuning and response properties. Our results demonstrate the practicality of using the Ising model to read out, or decode, spatial patterns of activity comprised of many hundreds of neurons.
\end{abstract}

\section{Introduction}

Interpreting the patterns of activity fired by populations of neurons is one of the central challenges of modern systems neuroscience. The design of decoding algorithms capable of millisecond-by-millisecond readout of sensory or behavioural correlates of neuronal activity patterns would be a valuable step in this direction. Such decoding algorithms, as well as helping us to understand the neural code, may have further practical application, as the basis of communication neural prostheses for severely disabled patients such as those with "Locked In" syndrome.

At the heart of such a decoding algorithm must lie - whether explicit or implicit - a description of the conditional probability distribution of activity patterns given stimuli or behaviours. Making this description is nontrivial, as the brain, like other biological systems, exhibits enormous complexity. This results in a very large number of possible states or configurations exhibited by the system, making the description of such systems by simply measuring the probabilities of each state unfeasible. Except for very small patterns, a model-based approach of some kind is essential.

New technologies in neuroscience such as high-density multi-electrode array recording and multi-photon calcium imaging now make it possible to monitor the activity of large numbers of neurons simultaneously. Analysis tools for such high dimensional data have however lagged behind the experimental technology, as most approaches are limited to very small population sizes. While considerable advances have been made in the use of information-theoretic approaches to characterise the statistical structure of small neural ensembles~\cite{Gawne1996,Panzeri1999,Schultz2001,Panzeri2001nc,Reich2001,Petersen2001,Pola2003,Montani2007}, finite sampling limitations have made results for larger ensembles much more difficult to obtain.

For the statistical description of multivariate neural spike train data, parametric models able to capture most of the interesting features of real data while still being of empirically accessible dimensionality are highly desirable. One promising approach has emerged from statistical mechanics: the use of Ising (or Ising-like) models, exploiting an analogy between populations of spike trains and ensembles of interacting magnetic spins \cite{Shlens2006,Shlens2009,Schneidman2006}.

Our aim here is to devise an algorithm for ``millisecond-by-millisecond" neural decoding, on the basis that information processing in the nervous system appears to make use of such fine temporal scales \cite{Carr1993,Bair1996}. The timescale of the ``symbols" used in information processing is thus likely to be somewhere between 1 and 20 ms for most purposes \cite{Butts2007}.
For time bins on this scale, neural spike trains are effectively binarized, and the simplest binary model (in the maximum entropy sense) that captures pairwise correlations is the Ising model.
The Ising model is thus a natural way to describe the statistics of neural spike patterns at the timescale of interest. Fitting of such a model to the observed neural data has the advantage that it does not implicitly assume some non-measured structure in the data, i.e. maximum entropy models express the most uncertainty about the modelled data given the (explicit) chosen constraints (e.g. that certain moments of the measured distribution agree with the model distribution) \cite{Jaynes1957}. It can be shown that this is mathematically equivalent to maximizing the likelihood of the model parameters to explain the observed data \cite{Berger1996}. By using this approach to fit a model to the conditional activity pattern distribution, in conjunction with maximum a posteriori decoding \cite{Foldiak93,Oram1998}, it is possible to train a decoder which takes as its input a pattern of spiking activity, and gives as its output the stimulus that it determines to have elicited that spike pattern.

A major obstacle to the use of Ising models for neural decoding, is that, in general, it is necessary to compute a partition function (or normalization factor), involving a sum over all possible states. This can be numerically challenging, and for large numbers of neurons, unfeasible. In the present study, we adopted several approaches for circumventing this problem. Firstly, we make use of mean field approximations, including both the `naive' mean field approximation and the \citeasnoun{Thouless1977} (TAP) extension to it, following \citeasnoun{Roudi2009b}. Secondly, we compare this with the recently proposed Minimum Probability Flow Method \cite{Sohl-Dickstein2010} for learning model parameters. To assess the relative performance of these approaches in the context of a discrete decoding problem, we simulated the activity of a population of neurons in layer V of the mouse visual cortex during an experiment in which a discrete set of orientation stimuli were presented. Using this simulation, we evaluated the relative performance characteristics of the different decoding algorithms in the face of limited data, exploring decoding regimes with up to 1000 neurons. We demonstrate, for the first time, the use of the Ising model to effectively decode discrete stimulus states from large-scale simulated neural population activity.

\section{Methods}

\subsection{Ising models of neural spike trains}

Activity states in an Ising model are Boltzmann distributed, i.e. they are distributed according to the negative exponential of the ``energy'' associated with each state.
This distribution, \\ $P \propto e^{-\sum_\mu \lambda_\mu f_\mu}$, is the maximum entropy distribution subject to the set of constraints imposed by Lagrange multiplers $\lambda_\mu$ on variables $f_\mu$.
Imposing these constraints upon firing rates and pairwise correlations gives
 \begin{equation}
 p_{\text{Ising}}(\mathbf{r}|s) = \frac{1}{\mathcal Z(s)} \exp \left( \sum_i h_i(s) r_i + \frac{1}{2}\sum_{i\neq j} J_{ij}(s) r_i r_j  \right),
\end{equation}
where $\mathbf{r} = (r_1,r_2,\ldots,r_C)^T$ and each binary response variable $r_i \in\lbrace0,1\rbrace$ indicates the firing/not firing of neuron $i$ in the observed time interval. The parameters $h_i$ (known in statistical physics as `external fields') and $J_{ij}$ (`pairwise couplings') have to be fit to the data such that the model displays the same means and pairwise correlations as the data:
\begin{subequations}
\begin{align}
 \left \langle r_i\right \rangle_{\text{Ising}} {=} & \left \langle r_i\right \rangle_{\text{Data}},\label{constraints1}\\
\left \langle r_ir_j\right \rangle_{\text{Ising}}{=} & \left \langle r_ir_j\right \rangle_{\text{Data}},
\end{align}
\end{subequations}
where $\langle \cdot\rangle_{\text{model}}$ denotes expectation with respect to the specified distribution.
$\mathcal Z(s)$ is the partition function, which acts as a normalisation factor. i.e.:
\begin{equation}\label{part_fc}
 \mathcal Z(s) = \sum_{\mathbf{r} \in \mathcal R} \exp \left( \sum_i h_i(s) r_i + \frac{1}{2}\sum_{i\neq j} J_{ij}(s) r_i r_j  \right) .
\end{equation}
Note that the first sum is over all {\em possible} (as opposed to observed) responses, given by the set $\mathcal R$.

In statistical physics it is more common to use a symmetric representation $\sigma_i \in \{-1,1\}$ for the `spins' that describe the activation of neuron $i$ (with $-1$ indicating `no spike' and $1$ indicating `spike'), which simply corresponds to a change of variables $\sigma_i = 2r_i -1$. Accordingly the fields, couplings and partition functions change. As it is occasionally more convenient to work in one or the other representation we will denote the fields and couplings in the spin representation with $\tilde h_i$ and $\tilde J_{ij}$.

Standard Monte Carlo techniques for fitting these model parameters, such as Boltzmann learning, which can in principle provide an exact solution - given the number of samples is high enough - become computationally very expensive if not intractable as the number of cells increases. We have found in previous work \cite{Seiler2009}, that the Boltzmann learning approach becomes computationally too expensive in our case for ensemble sizes larger than 30 cells.
This poor scaling behaviour is mainly due to the exponentially increasing number of states with the number of cells.
Speeding up the model fitting process is hence an essential requirement to utilize Ising models for studies with large ensembles of neurons.
Solutions to speed up the ``classical'' Boltzmann learning approach have been suggested \cite{Broderick2007}. However these are still associated with a high computational cost.

\subsection{Neural Decoding}
In this paper we consider the problem of decoding which of a number of different stimuli has elicited a neural spike pattern. This can be seen as a discrete classification task: we have a set of $S$ stimuli $s \in \mathcal S = \{s_1,s_2,\ldots,s_S\}$. Decoding in this scenario means that we have to provide a decision rule that estimates which stimulus has been the input to the system, given an observed spike pattern $\mathbf{r}_{\text{obs}}$. The particular example to which we apply this is a simulation of the spike pattern responses elicited by visual stimuli across the receptive field of a visual cortical neuron: in this case each stimulus $s_i$ represents a different orientation of a sinusoidal grating. Our main aim with this simulation was to validate our methodology in a neurophysiologically realistic coding regime, relevant to datasets to which our methodology might be applied. As a supplementary goal, we hoped to gain some insight into whether some aspects of the model affect decoding performance - such as heterogeneity of tuning, observed in real neural recordings but often ignored in population coding models.

For decoding, we use the maximum a posteriori (MAP) rule:
\begin{eqnarray}
 \hat s &=& \arg \max_{s}  p(s|\mathbf{r}_{\text{obs}}) \\
 &=& \arg \max_{s} \frac{p(\mathbf{r}_\text{obs}|s)p(s)}{p(\mathbf{r}_\text{obs})} \\
 \null&=& \arg \max_{s} p(\mathbf{r}_\text{obs}|s)p(s),
\end{eqnarray}
where the second step is the application of Bayes' theorem and the third equality holds because $p(\mathbf{r}_\text{obs})$ is independent of $s$ and is hence irrelevant for maximising the given expression, i.e. just a constant factor with respect to $s$ that scales the maximum accordingly. In the case we examine here we assume we are in control of the stimulus distribution $p(s)$, and thus we can choose it to be uniform, i.e. to exhibit the same constant probability for each stimulus and therefore be independent of $s$, as well. Hence our decoding rule simplifies further to the maximum likelihood (ML) rule:
\begin{equation}\label{ML_rule}
 \hat s = \arg \max_{s} p(\mathbf{r}_\text{obs}|s).
\end{equation}
Within this setting, the task of creating a neural decoder reduces to the modelling of the stimulus dependent distributions $p(\mathbf{r}|s)$. Once these are obtained, we can apply our ML decoding rule (Equation \ref{ML_rule}) to estimate the given input stimulus $s$.

We have used two different statistical models to fit the observed spike patterns for each stimulus.
Firstly, we have used an Ising model for $p(\mathbf{r}|s)$, i.e. we assume that for each stimulus, the spike pattern distribution can be described by a (different) Ising model.
Secondly, we have used an independent model distribution $p_{\text{ind}}$, assuming that given a stimulus, each cell is independent of the others:
\begin{equation}
 p_{\text{ind}}(\mathbf{r}|s) = \prod_{i=1}^C p(r_i|s).
\end{equation}
The independent model is the binary maximum entropy model of first order, i.e. it takes into account only the first order moments (the constraints on the means given by Equation \ref{constraints1}) and is therefore a natural comparison for the Ising model. As it is very easy to fit the independent model, we used this as a control method, to test whether the more complex Ising model could enhance decoding performance. Note that the numerical values for the probabilities can get very small for large cell ensembles, and therefore to evade finite precision problems we use in this case an equivalent log-likelihood decoding rule instead of the ML rule, i.e. maximise the logarithm of the likelihood instead of maximising the likelihood directly.

\subsection{Training and Testing}
To train and test the decoders, we proceed as follows:
\begin{enumerate}
 \item For each stimulus we simulate a set of possible response vectors. The details of the simulation are described in the following subsection.
 \item We separate the simulated response patterns into training data, which is used to fit the model and test data which we use to evaluate the decoding performance of the obtained models.
 \item The whole testing procedure is performed with 10 fold cross-validation, i.e. we divide the whole data for each stimulus into 10 equally sized parts. We then use 9 parts of the data to train our model and the remaining one for testing. We repeat this process again with all 10 possible test/training data set combinations of this kind to reveal if our results generalize to the whole dataset.
\end{enumerate}

\begin{table*}[htb!]
 \centering
 \begin{tabular}{lcp{9.5cm}}
\toprule
\textbf{Parameter} & \textbf{Value} & \textbf{Comments}\\
\midrule
Preferred tuning direction & uniformly spaced $0^\circ\!-\!360^\circ$& \\
Tuning width (HWHM) & $38 + \xi\varPhi_i$  & $\xi \in [0,1]$; $\varPhi_i$ normally distributed, fitted to Fig. 4f of \citeasnoun{Niell2008}, truncated minimum of 10. \\
Direction Selectivity Index & $0.1$ & fixed for all neurons; \citeasnoun{Niell2008} Fig. 5c. \\
Spontaneous firing rate & $1.7 + \gamma X_i$ (Hz) &  $\gamma\in[0,1]$; $X_i$ normally distributed about zero, distribution fitted to Fig. 8d of \citeasnoun{Niell2008}; truncated to minimum of 0.3\\
Sustained evoked firing rate & $7 + \gamma Y_i$ (Hz) & $\gamma\in[0,1]$; $Y_i$ normally distributed, fitted to data in Fig. 8 and supp. fig. S3 of \citeasnoun{Niell2008}; truncated to minimum of zero\\
Transient-to-sustained ratio & $1.5$ & fixed for all neurons\\
\bottomrule
\end{tabular}
 \caption{Parameters of the mouse visual cortical model simulated. $\gamma$ and $\epsilon$ determine the extent of heterogeneity in the model, with $\gamma,\epsilon=0$ setting the model properties to homogeneous.}
 \label{tab:simsett}
\end{table*}

\subsection{Simulation of Evoked Spike Patterns from Mouse Primary Visual Cortex}

We simulated the transient response patterns of activity evoked by visual (orientation) stimuli in layer V pyramidal neurons of the anaesthetized mouse visual cortex.
The orientation direction were chosen to be $n\cdot180/S$, where S is the number of stimuli and $n \in \{0,1,\ldots,S-1\}$.
The properties of our simulation are motivated by the results reported by \citeasnoun{Niell2008}.
We simulated different models by augmenting a basic model with mostly homogeneous response characteristics, with some come controlled heterogeneous characteristics.

Our model is defined as follows, with parameters specified in Table~\ref{tab:simsett}.
The spontaneous activity of each neuron was set to 1.7 spikes per second, corresponds to the reported median value for layer V neurons in \cite{Niell2008}. We assumed that neuronal direction preferences were uniformly spaced around the circle. Each neuron's tuning curve was defined by a von Mises function (circular Gaussian) with half width at half maximum (HWHM) fit to experimental data \cite{Niell2008}. The direction selectivity index (DSI) was set to 0.1 for all layer 5 neurons in our model. Sustained firing rates were fit to the distributions reported in \cite{Niell2008}. To reflect that we are considering a situation in which a stimulus is decoded from a short time window (20 ms) of data, we multiplied these evoked rates by a fixed transient-to-sustained ratio of 1.5, taken to reflect the onset response of the neuron's response to a flashed stimulus. As our model is fit to data from directional (drifting grating) stimuli, we took the arithmetic mean value of the two corresponding diametrically opposite directions for each neuron to compute the model response to a flashed orientation.

\begin{figure*}[tb]
 \centering
 \includegraphics{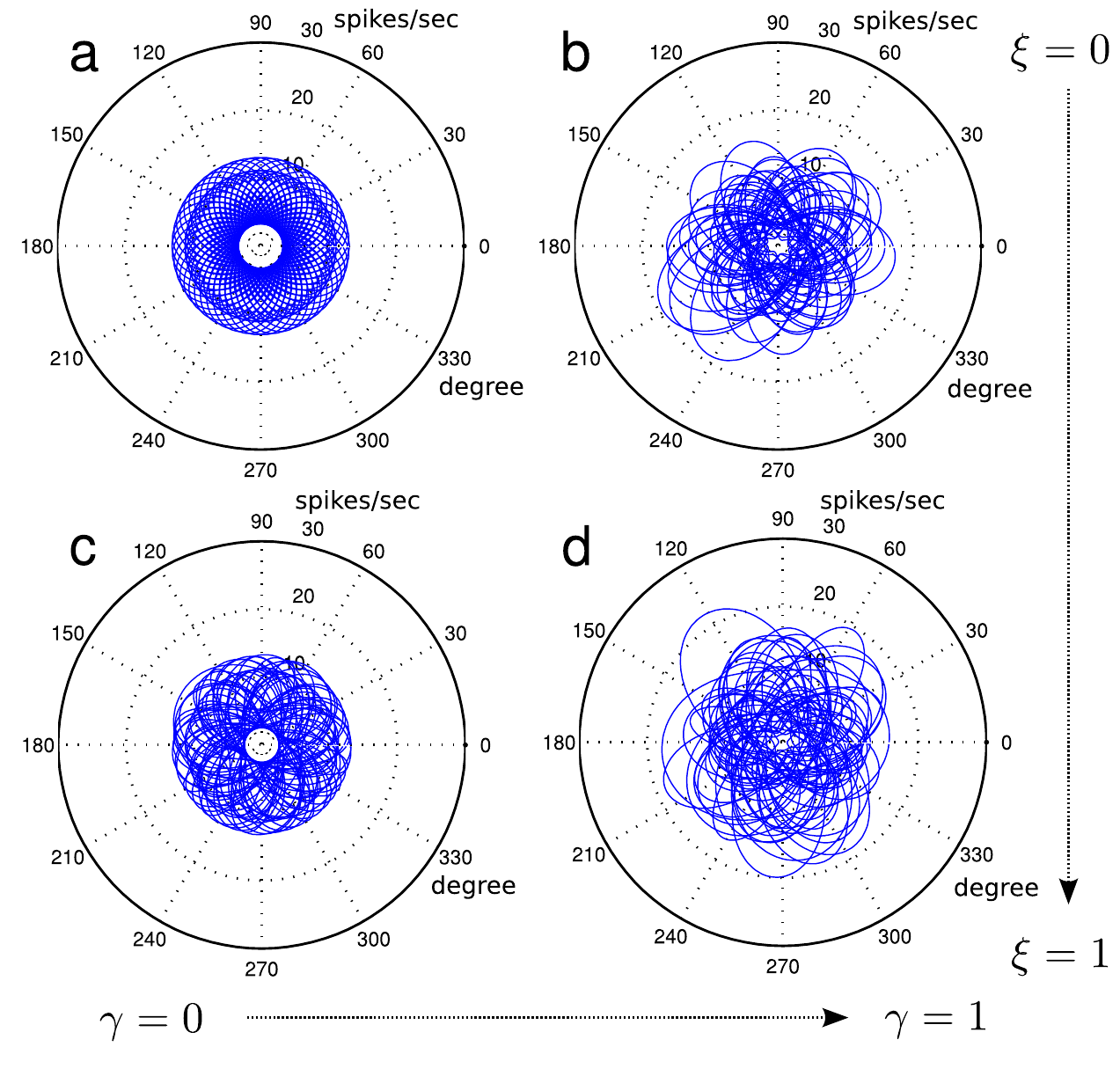}
 \caption{Tuning curves for a neural ensemble of 50 cells for different heterogeneity parameters $\gamma,\xi$. The shown spiking rates correspond to the transient rates. \textbf{a} Tuning curves for basic model, $\gamma = \xi = 0$. \textbf{b} Tuning curves with heterogeneous firing rates ($\gamma = 1$), but fixed tuning widths ($\xi = 0$) (NB the tuning width is defined relative to the spontaneous activity). \textbf{c} Tuning curves with heterogeneous tuning widths ($\xi=1$), but fixed firing rates ($\gamma =0$). \textbf{d} Tuning curves for fully heterogeneous scenario ($\gamma = \xi = 1$)}
 \label{fig:2}
\end{figure*}

The characteristics of the basic model were modulated via two inhomogeneity parameters $\gamma,\xi\in[0,1]$, to introduce a heterogeneous distribution of firing rates and the tuning widths respectively, as described in Table~\ref{tab:simsett}. We neglected inhomogeneity in other parameters. Thus the parameter $\gamma$ regulates firing rate heterogeneity, where $\gamma = 0$ corresponded to the basic homogeneous mode,l and $\gamma=1$ to the most heterogeneous firing rate setting. The parameter $\xi$ was used analogously to regulate heterogeneity in  the tuning widths of the neurons. The effects of the two heterogeneity parameters $\gamma, \xi$ are illustrated in Figure \ref{fig:2}.

Patterns of spikes fired by the neural population were simulated using a dichotomized Gaussian approach \cite{Macke2009}. Since we cannot estimate covariance matrices from experimental data directly, and not every positive definite symmetric matrix can be used as the covariance matrix of a multivariate binary distribution, we adapted the following approach.
First we compute upper and lower covariance bounds for each pair of neurons, according to \cite{Macke2009}
\begin{eqnarray}
\max \left\{-pq , -(1-p)(1-q)\right\}  \le \mathrm{Cov}(r_i, r_j) \nonumber \\
\le \min \left\{ (1-q)p, (1-p)q\right\},
\end{eqnarray}
where $p$ and $q$ are the means (mean spiking probabilities) of neuron $i$ and $j$, respectively.
We then choose a random symmetric matrix $A$ that lies between these bounds. As in general this choice does not result in a permissible correlation matrix for the underlying Gaussian, a \citeasnoun{Higham2002} correction is applied to find the closest correlation matrix possible for the latent Gaussian (cf. \citeasnoun{Macke2009}), to which we finally arithmetically add a random correlation matrix with uniformly distributed eigenvalues to adjust the mean correlation strength. Having established a dichotomized Gaussian model we can thus draw samples with high efficiency.

\begin{figure*}[tb]
 \centering
 \includegraphics{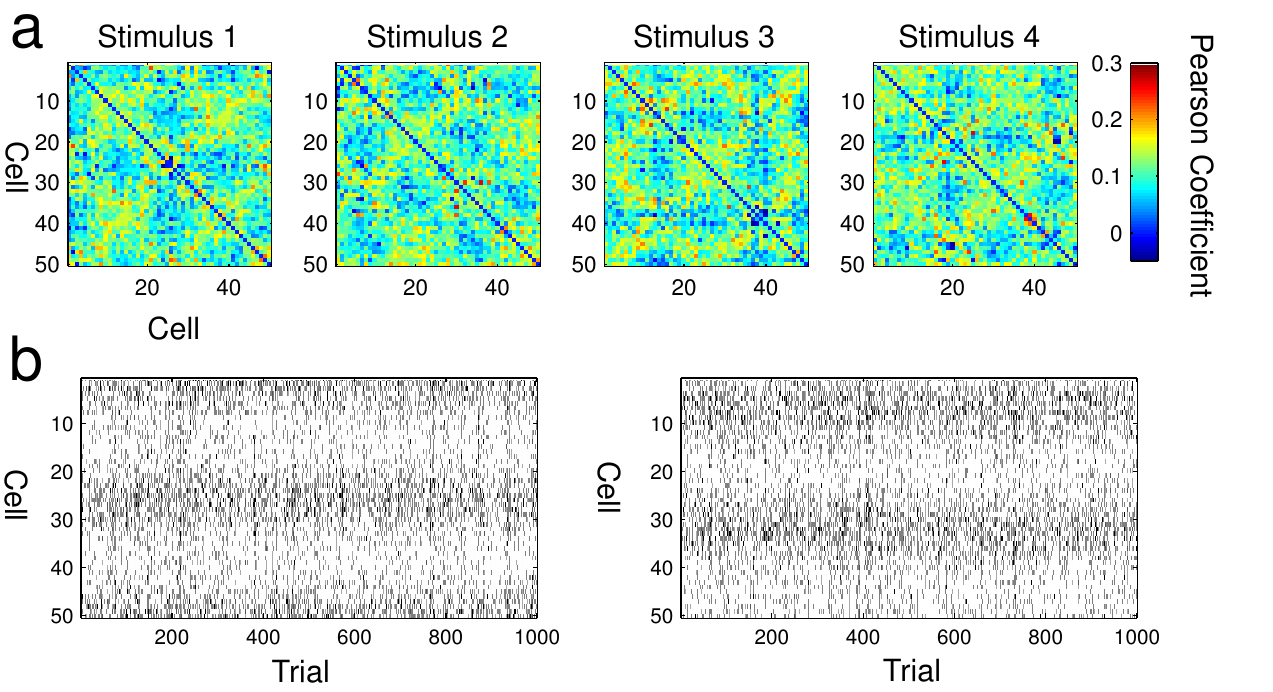}
 \caption{Neural ensemble responses simulated for basic model ($\gamma =\xi = 0$) for 50 cells. \textbf{a} Correlation matrices for 4 stimuli (computed with 100000 samples). Diagonal terms set to zero for visualization purposes only. \textbf{b} Simulated population neural responses over 1000 trials for two different stimuli, with black indicated a spiking response from a neuron on a given trial.}
 \label{fig:3}
\end{figure*}

Where not otherwise stated in the text, 10000 trials per stimulus were simulated, allowing 9000 training samples and 1000 test samples with 10-fold crossvalidation. In the absence of a detailed characterization of the correlation structure of neural responses in the mouse visual cortex, we assumed that the correlation in firing between each pair of neurons was weak and positive. Our simulation results in a mean correlation of 0.11 and a standard deviation of 0.040 (measured with 100000 samples for different ensemble sizes) for our basic model and similar levels of correlation for nonzero $\gamma,\xi$.
Due to limitations of the dichotomized Gaussian simulation, we were not able to specify the correlations of the spike trains/between the individual neurons exactly, thus all reported correlations are measured and may be prone to small variations. However, such limitations would be inherent to any simulation approach, as i) the covariance structure of a multivariate binary distribution is always constrained by the firing probabilities of the cells and can not be chosen independently of these firing rates \cite{Macke2009}, and ii) finite sampling effects will always affect the simulated data, resulting in fluctuations in the correlation structure.

We were able to vary the (measured) mean absolute correlation level in some simulations, allowing an assessment of the relative effects of correlation strength for decoding. To do this we proceeded as follows: having established the correlation matrix of the latent Gaussian as described before allows us to sample in a regime with correlations around $0.1$. Likewise a latent Gaussian with a correlation matrix given by the identity matrix would correspond to the case where there are no correlations in the latent Gaussian and thus in the simulated spike trains itself, which means that the simulated spike patterns for each neuron are independent (note that due to the model used in \citeasnoun{Macke2009} the correlation matrix and the covariance matrix of the latent Gaussian are the same).  Finite sampling effects might still introduce some nonzero measured correlations in the spike patterns, but the underlying distribution would still be with independent neurons. Thus, by interpolating between these two cases we can reduce the effective correlations in the data in a controlled way, where at one end we yield our original model and at the other end we yield a set of independent neurons.

\subsection{Fitting the Ising Model Parameters}
For fitting the model parameters in the Ising model case we use two different strategies: mean field approximations and minimum probability flow learning. In earlier work we used Boltzmann learning \cite{Seiler2009}, however as this becomes rapidly computationally intractable with an increasing number of neurons, we have not reported it here.

\subsubsection{Mean Field Methods}
The suitability of different Mean Field approaches for fitting the parameters of an Ising model have been recently assessed by \citeasnoun{Roudi2009b}. In their work, Roudi et al. compared the learned model parameter as inferred by Boltzmann learning, with the parameters inferred by a number of different approximative algorithms. Here we examine the utility for decoding of using successively higher order mean field approximations. Tanaka \citeyear{Tanaka1998} demonstrated how to systematically obtain mean field approximations of increasing order based on a Plefka series expansion \cite{Plefka2006} of the Gibbs free energy. By truncating these series to terms up to n-th order, and using the linear response correction, it is possible to derive an n-th order mean field approximation, yielding $C$ equations for the external fields, and $C(C-1)/2$ equations for the pairwise couplings (cf. \citeasnoun{Tanaka1998}). These equations can then be solved with respect to $\tilde J_{ij}$ and $\tilde h_i$. For higher order approximations, these equations can have more than one solution. This problem can be resolved by considering that the Plefka series expansion is effectively a Taylor series expansion and continuity of the solutions is expected when higher order terms are gradually increased \cite{Tanaka1998}.

Tanaka further provided an explanation for the ``diagonal weight trick'' as used by \citeasnoun{Kappen1998}. With this trick one introduces $C$ extra equations for the pairwise ``self-couplings'' $\tilde J_{ii}$, which can be used to refine the respective approximations. The success of this trick can be explained \cite{Tanaka1998} by considering that using this diagonal weight trick in an n-th order method, is effectively incorporating the dominant terms of the next higher $(n+1)$ order expansion.

For fitting the parameters, we have compared different mean field approximations from zeroth to second order with and without the diagonal weight trick, thus incorporating with our highest approximations all second order terms and the dominant third order terms of the free energy expansion. The zeroth order method is thereby equivalent to the independent model, thus providing another way of thinking about the independent decoder.

\paragraph{First order methods (naive mean field approximation).}
For the first order or naive mean field approximation, the equations for the external fields and pairwise couplings become:
\begin{subequations}
 \begin{align}
  &\mathbf{\tilde J} = - \mathbf{C^{-1}} \qquad \qquad(i \neq j),\\
  &\tilde{h}_i = \tanh^{-1}m_i - \sum_{j\neq i} \tilde{J}_{ij} m_j,
 \end{align}
\end{subequations}
where $\mathbf{\tilde J}$ is the estimated coupling matrix with elements $\tilde{J}_{ij} $, the `magnetization'  $m_i =\langle\sigma_i\rangle$, and the covariance matrix $\mathbf{C}$ is defined by $C_{ij} = \langle\sigma_i \sigma_j\rangle -m_im_j$. In the following we will denote this naive mean field method by nMF.

By incorporating the diagonal weight trick the above equations change slightly into the following:
\begin{subequations}
\begin{align}
& \mathbf{\tilde J} = \mathbf{P}^{-1} - \mathbf{C}^{-1},&\\
 &\tilde{h}_i = \tanh^{-1}m_i - \sum_{j} \tilde{J}_{ij} m_j,&
\end{align}
\end{subequations}
where in addition to the matrices defined above, we have defined the diagonal matrix $P_{ij} = (1-m_i^2)\delta_{ij}$, with $\delta_{ij}$ being the Kronecker delta.
Note that in the nomenclature of \citeasnoun{Roudi2009b}, this has been called ``naive mean field method''. However we will in the following refer to it as naive mean field method with diagonal weight trick (nMFwd).

\paragraph{Second order methods (TAP approximation).}
The equations of the second order method are also known as the TAP equations \cite{Thouless1977} and can be seen as a correction to the naive mean field methods. Using the TAP approach the equations for the model parameters read:
\begin{subequations}
 \begin{align}
  &2\tilde{J}_{ij}^{\:2}m_im_j + \tilde{J}_{ij} + (\mathbf{C}^{-1})_{ij}=0 \qquad(i\neq j),\\
  &\tilde{h}_i = \tanh^{-1}m_i - \sum_{j\neq i} \tilde{J}_{ij}m_j +m_i\sum_{j \neq i}\tilde{J}_{ij}^{\:2}(1-m_j^2),
 \end{align}
\end{subequations}
where the first equation can be solved for the pairwise couplings $J_{ij}$. As mentioned previously the correct solution has to be chosen according to continuity conditions outlined in \citeasnoun{Tanaka1998}, from which then the external fields $h_i$ can be computed.
More precisely, if $m_im_j(\mathbf{C}^{-1})_{ij}<0$ we choose the solution, which is closer to the original first order mean field solution. If $m_im_j(\mathbf{C}^{-1})_{ij}>0$ we use the first order mean field solution directly.
We use this procedure as it avoids pairwise couplings becoming complex, and respects the continuity of the inverse Ising problem for $J_{ij}$ as a function of $(\mathbf{C}^{-1})_{ij}$. In the following this method is denoted by TAP.

\paragraph{Incorporating third order terms.}
The second order method can also be augmented by the diagonal weight trick, hence incorporating the leading third order terms of the free energy expansion. The equations for the TAP method with diagonal weight trick (TAPwd) are given by:
\begin{subequations}
 \begin{align}
  &2\tilde{J}_{ij}^{\:2}m_im_j + \tilde{J}_{ij} + (\mathbf{C}^{-1})_{ij}=0 \qquad(i\neq j),\\
   &J_{ii} =\dfrac{1}{1-m_i^2} - (\mathbf{C}^{-1})_{ii} \qquad (i=j),\\
 &\tilde{h}_i = \tanh^{-1}m_i - \sum_{j} \tilde{J}_{ij}m_j,
 \end{align}
\end{subequations}
where we have solved the equations in an analogous fashion to that described for the normal TAP approach.

\subsubsection{Minimum Probability Flow Learning}
\citeasnoun{Sohl-Dickstein2010} recently proposed the Minimum Probability Flow Learning (MPFL) technique, which provides a general framework for learning model parameters. As this technique is also applicable to the Ising model, we have used it to learn external fields and pairwise couplings for our model. However, as the sampling regime usually feasible in neurophysiological experiments dictates a small number of samples compared to the number of parameters in the model (which is $\mathcal O(C^2)$ with $C$ cells), the learning problem for the parameters becomes under-constrained already at intermediate neural ensemble sizes, i.e. we are likely to have more parameters to fit than there are samples.

We therefore introduced a regularization term to their original objective function to penalize model parameters growing to large numbers, i.e. to avoid overfitting. Given the original objective function $K(\theta)$ with $\theta$ being the parameters of our model, our regularized objective function reads:
\begin{equation}
 K_{\text{reg}}(\theta) = K(\theta) + \frac{\lambda}{2} \| \theta \|_{2}^2,
\end{equation}
where $\|\cdot\|_2$ is the $L_2$ norm, which is a common choice of regularization term \cite{Bishop2007}. So for the Ising model case we have:
\begin{equation*}
\|\theta\|_2^2 = \sum_{i} h_i^2 + \sum_{i\neq j} J_{ij}^2.
\end{equation*}
For the present work, the regularization parameter was set to $\lambda = 0.0127$, after systematically assessing different settings for an ensemble size of $100$ cells via cross-validation. We refer to this learning algorithm as rMPFL (regularized MPFL) in the rest of the manuscript.

Other choices for the regularization term are possible and might even result in better performance for decoding purposes, e.g.  two independent penalty terms for the external fields and the pairwise couplings. However an extensive assessment of different parameter settings would be very time consuming due to the cost of calculating the invoked partition function.  We therefore have not performed an exhaustive analysis of regularization. In a real experiment the regularization term should however be adapted to yield the best possible performance for the specific number of cells. We not that while it is not necessary to compute the partition function for some applications of MPFL (e.g. if learning the $J_{ij}$ parameters is the end in itself), it is required for decoding.

\subsection{Partition Function Estimation}
Estimating the partition function is a computationally expensive task, since the set of possible responses $\mathcal R$ grows exponentially with the number of cells $C$, rendering an analytical computation (Equation \ref{part_fc}) intractable for large neural ensembles.

As MPFL learning does not provide an estimate of the partition function, we used the Ogata-Tanemura partition function estimator \cite{Ogata1984,Huang2001}, which is based on Markov Chain Monte Carlo (MCMC) techniques. However MCMC is still a very time consuming technique. To speed up the model fitting process, we estimated the partition function for each stimulus only once in a 10 fold cross-validation run when using MPFL, as ideally all samples for a specific stimulus should come from the same distribution, thus approximately sharing the same partition function. We have examined the effect of this approximation (see Results).

When fitting the model parameters with mean field theoretic approaches, we computed the (true) partition function $\mathcal Z (s)$ in the mean field approximation, as reported in \citeasnoun{Kappen1998}, \citeasnoun{Thouless1977}, and \citeasnoun{Tanaka1998}.

For the first order mean field approach this yields:
\begin{equation}
\begin{split}
 &\log \mathcal Z = \sum_i \log\left(2\cosh\left(\tilde h_i +  W_i\right)\right) - \sum_i W_im_i \\
 &\qquad \quad+\sum_{i<j}\tilde J_{ij}m_im_j,\\
\end{split}
\end{equation}
with
\begin{equation*}
 W_i = \sum_{j\neq i} \tilde J_{ij}m_j.
\end{equation*}
Here each of the parameters is actually a function of the stimulus $s$, which we omit for clarity.

For the second order methods, the corresponding equation becomes:
\begin{equation}
\begin{split}
 &\log \mathcal Z = \sum_i \log\left(2\cosh\left(\tilde h_i + L_i\right)\right) - \sum_i L_im_i\\
 &\quad \quad +\sum_{i<j}\tilde J_{ij}m_im_j + \frac{1}{2}\sum_{i<j} \tilde J_{ij}^{\:2}(1-m_i^2)(1-m_j^2),
\end{split}
\end{equation}
with
\begin{equation*}
 L_i = \sum_{j\neq i} \tilde J_{ij}m_j - m_i \sum_{j\neq i}\tilde J_{ij}^{\:2}(1-m_j^2).
\end{equation*}

\subsection{Performance evaluation}

The fraction of correctly decoded trials was the principal method used to assess decoding performance. However, as the fraction correct or accuracy does not by itself provide a complete description of decoder performance, we sought used additional performance measures. The performance of a decoder is fully described by its confusion matrix, and we show how directly examining this matrix can yield insight into its behaviour. However, it is advantageous to be able to reduce this to a single number in many cirumstances. We therefore additionally computed the mutual information between the encoded and decoded stimulus \cite{Panzeri1999dec} to characterise the performance further. This provides a compact summary of the information content of the decoding confusion matrix.

We can write the mutual information (measured in bits) as:
\begin{equation}
 I(s,\hat s) = H(s) - H(s|\hat s),
\label{eq:mutinfo}
\end{equation}
where $H(s)$ is the entropy of the encoded stimulus
\begin{equation}
 H(s) = -\sum_{s\in\mathcal S} p(s) \log_2 p(s),
\end{equation}
and $H(s|\hat s)$ is the conditional entropy describing the distribution of stimuli $s$ that have been observed to elicit each decoded state $\hat s$,
\begin{equation}
 H(s|\hat s) = -\sum_{s,\hat s \in \mathcal S} p(s,\hat s) \log_2 p(s|\hat s).
\end{equation}
Since we have in the current study opted for a uniform stimulus distribution, the entropy $H(s)$ is simply given by
\begin{equation}
 H(s) = \log_2 S.
\end{equation}
In general the conditional entropy $H(s|\hat s)$ has to be computed from the confusion matrix. We note that if we were to assume that the correctly decoded stimuli and errors are uniformly distributed for all stimuli, i.e. that the conditional distribution $p(\hat s|s)$ is of the form
\begin{equation*}
 p(\hat s|s) =
\begin{cases}
f_c & \text{for } \hat s = s\\
\dfrac{1-f_c}{S-1}& \text{for } \hat s \neq s ,
\end{cases}
\end{equation*}
then the conditional entropy simplifies to
\begin{equation*}
 H(s|\hat s) = - f_c\log_2 f_c - (1-f_c)\log_2\left(\frac{1-f_c}{S-1}\right) .
\end{equation*}
This simplified expression has been used to characterize decoder performance in the Brain Computer Interface literature \cite{Wolpaw2002}. Here, we present this equation only to make apparent the scaling behaviour, and compute the decoded information using the more general expression (Eqn.~\ref{eq:mutinfo}).

\section{Results}

We performed computer simulations as described in Methods, to generate datasets for training and testing decoding algorithms. For all settings, 20 simulations were performed with different random number seeds, in order to characterize decoding performance. A number of metrics, including the fraction of correct decodings (accuracy) and mutual information between decoded and presented stimulus distributions, were used in order to characterize and compare decoding performance.

\begin{figure*}[tb]
 \centering
 \includegraphics{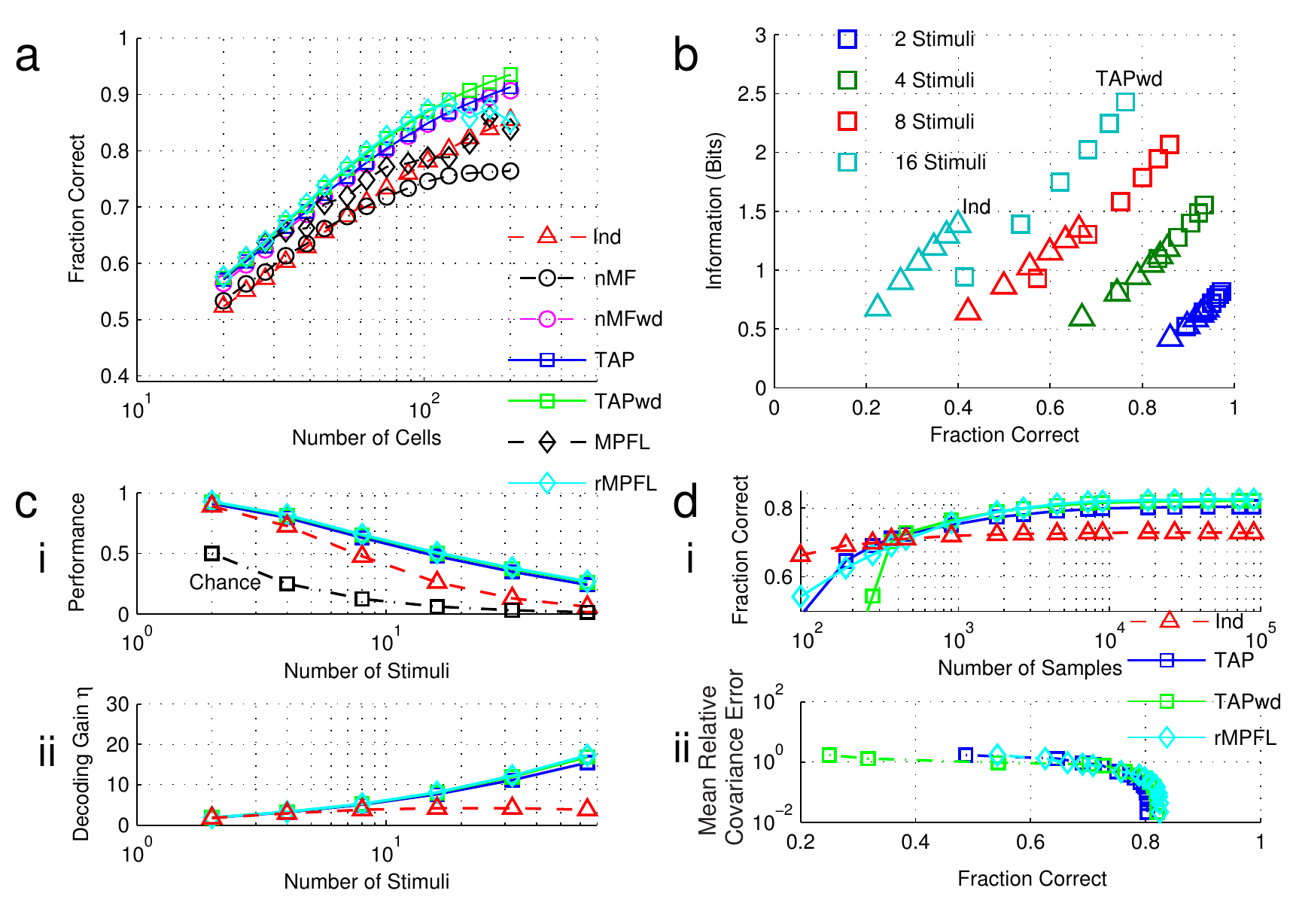}
 \caption{Performance of decoding algorithms for the basic model. \textbf{a} Fraction of correct decodings versus neural ensemble size, for a training dataset size of 9000 samples per stimulus. \textbf{b} The relationship between fraction correct and mutual information $I(s,\hat s)$ for varying stimulus set and ensemble size  ($C=\left\lbrace 50, 80, 110, 140, 170, 200\right\rbrace $ varying from bottom left to top right for each symbol type). Triangles denote the performance of the Independent decoder, squares the TAP Ising Decoder with diagonal weight trick. \textbf{c} The dependence of decoding performance on stimulus set size for 70 cells. \textbf{i} TAP, TAPwd, Independent and rMPFL decoders compared to random selection of stimuli. This is replotted in \textbf{ii} as the gain in fraction correct over chance performance, $\eta = p_\text{dec}/p_\text{guess}$, making the performance saturation for the independent decoder as problem difficulty increases more apparent. \textbf{d} Dependence of decoder performance on training set sample size, for 4 stimuli and a neural ensemble size of 70. \textbf{i} Fraction correct as a function of number of training samples. Below 450 training samples the Ising decoder fails to better the independent decoder.  \textbf{ii} Relation between mean relative covariance error $E$ as a measure of finite sampling effects and fraction correct.}
 \label{fig:4}
\end{figure*}

\subsubsection*{Basic Model}

The Ising model based decoders show better performance than the independent decoder in nearly all cases (Fig.~\ref{fig:4}a), in terms of the average fraction of correctly decoded stimuli. The performance of the standard (non-regularized) MPFL technique, however, in our hands falls away relatively quickly as the number of cells increases, failing to better the independent decoder after approximately 100 cells. This behaviour can be explained by considering that the problem of parameter estimation becomes more and more underconstrained as the number of cells increases while holding the number of training samples fixed. Falsely learned model parameters moreover affect the decoding performance by influencing the estimated partition function and thus worsen the decoder performance. As we have only estimated the partition function once per stimulus when using MPFL, large fluctuations in the training dataset can potentially have a big effect. To compensate for this behaviour, a regularization term can be included, which can stabilize the performance up to a significantly larger number of neurons (as described in Methods). Our regularized version of MPFL still decreases in performance after about 110 cells. However, we have used a fixed value for the regularization parameter for all simulations here, whereas ideally the regularization term should be adapted to the number of cells and number of samples in the specific setting. Using such an approach would most likely result in a better performance for larger numbers of cells. As an example we have tested for an optimized $\lambda$ parameter for 150 cells and found that we could increase performance to $0.874$ in this case (average over 10 trials, not shown in figure).
\begin{figure}[tb]
 \centering
 \includegraphics{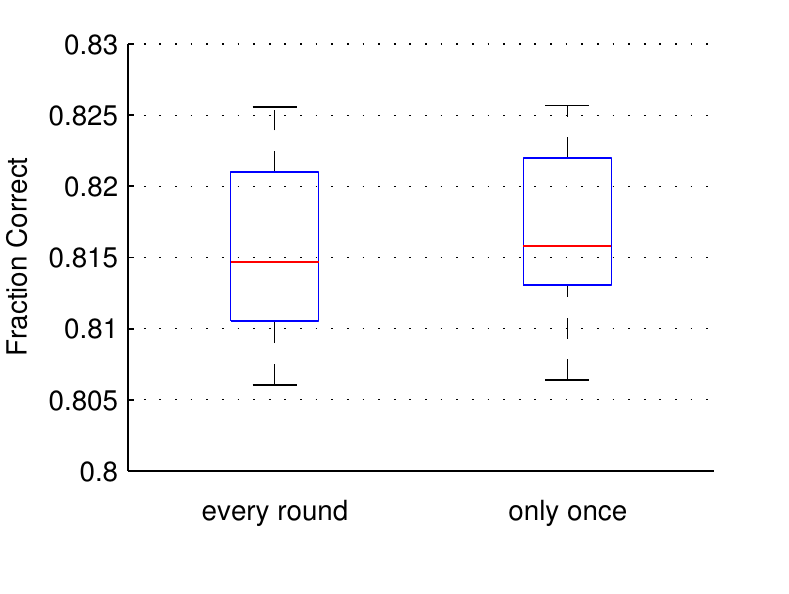}
 \caption{Decoding performance effects of reduced partition function estimation for rMPFL method as Box-Whisker plot. Interquartile range is indicated by blue box, the median value by a red line. Whiskers are used to visualize the spread of the remaining data for values at most within 1.5 times the interquartile range from the ends of the box. Left: results for computing the partitioning function for each stimulus every round for 10-fold crossvalidation. Right: results for computing the partition function only in the first round for each stimulus. Data from 20 simulation runs, 70 cells, basic model, 10000 samples simulated per stimulus}
 \label{fig:5}
\end{figure}

One of the assumptions made for much of this paper is that the partition function can be estimated only once in each 10-fold crossvalidation run, thus speeding up training. We studied the effect of this approximation on performance, finding, as shown in Figure \ref{fig:5}, that the effect is marginal, at least under our operating conditions. No statistically significant difference between the two approaches was observed. In a decoding regime where the rMPFL method starts to infer the wrong model parameters (e.g. due to overfitting) one would expect these effects to become more pronounced. However, in such a scenario where the rMPFL method becomes unreliable, a different regularization term or a different inference method should be considered.

The decoded information analysis reveals that the difference in the decoding performance of the independent and Ising (as exemplified by TAP, TAPwd and rMPFL) models becomes more pronounced as the number of stimuli is increased, as shown in Fig.~\ref{fig:4}b,c. As the number of stimuli increase, the independent and Ising decoder curves separate, indicating not only a difference in the accuracy of both decoders, but also a difference in the confusion matrices, i.e. in the distribution of errors between the two approaches. This is considered in more detail in section~\ref{sec:confmat}.

A further interesting behavior of the Ising decoder is apparent in Fig.~ \ref{fig:4}c: as the number of stimuli increases, the relative decoding gain $\eta$, (which we define as the ratio between the ``actual" and ``chance" fraction correct) keeps increasing with the number of stimuli for the Ising model case, whereas it saturates for the independent decoder. This effect can be explained as follows: the independent decoder relies on the fact that two different stimuli will result in different firing rates for each cell. With an increasing number of stimuli however, the difference between two adjacent stimuli becomes smaller and smaller, and thus the difference in the firing rates discriminating two adjacent stimuli becomes smaller and smaller. Therefore the decoder performance of the independent decoder rapidly decreases as the number of stimuli increases. As two adjacent stimuli can result in neural responses that have quite different correlation structure despite having very close firing rates, the Ising decoder can additionally make use of this information in the data and thus yield better performance.
This suggests that the Ising decoder may be particularly advantageous as the decoding problem becomes more difficult and not easily discriminable in terms of the neural firing rates. Real world performance will of course be dependent upon the level of systematic difference in correlation structure induced by stimuli in any given dataset.

Performance of the Ising decoder is strongly dependent on the number of training trials available (Fig.~ \ref{fig:4}d i). Here we found, for 70 neurons, that around 400 training samples were required to allow the Ising decoder to outperform independent decoding. The independent decoder will necessarily have better sampling performance, as it relies only upon lower order response statistics. (It is worth recalling that a ``full" decoder, which made use of all aspects of spike pattern structure, would have far worse scaling behaviour than either). Another way of looking at this is to examine the estimated covariance matrix from the simulated data (Fig.~ \ref{fig:4}d ii).
We define the mean relative error of the covariance matrix as
\begin{equation}
 E = \dfrac{1}{S}\sum_{s\in\mathcal S}\dfrac{1}{C^2}\sum_{i,j} \left|\dfrac{C_{ij,s}^\text{data}-C_{ij,s}^\text{as}}{C_{ij,s}^\text{as}}\right|,
\end{equation}
where $C_{ij,s}^\text{data}$ is the estimated covariance between unit $i$ and $j$ under stimulus $s$ from the (finite) simulated data (normally 10000 samples), and $C_{ij,s}^\text{as}$ is the asymptotic covariance, defined in the same way but computed with 100,000 samples. This error provides a measure of the finite sampling bias we encounter for fitting the model.

\begin{figure}[tb]
 \centering
 \includegraphics{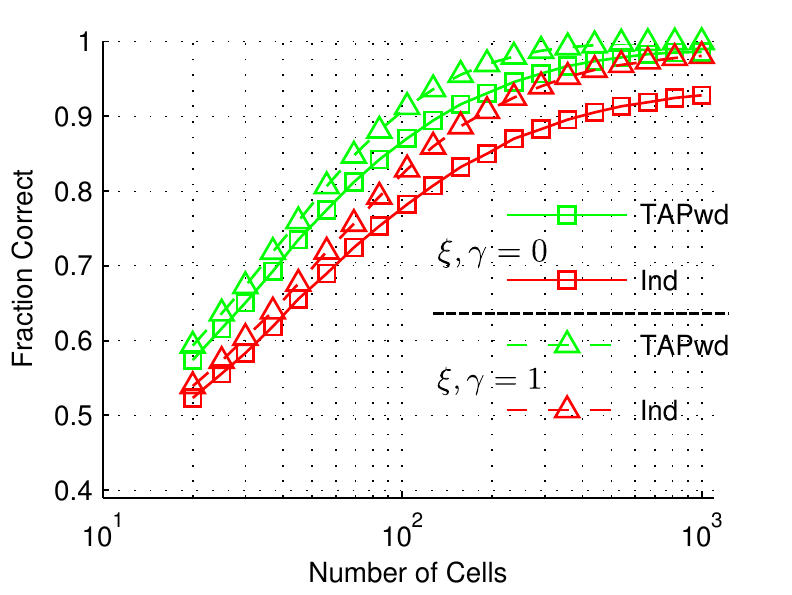}
 \caption{Decoder performance is enhanced in the heterogeneous scenario ($\xi=\gamma=1$). Comparison of Fraction correct vs. number of cells for basic model and fully heterogeneous model. The TAPwd algorithm was used to train the decoder.}
 \label{fig:6}
\end{figure}

\subsubsection*{Heterogeneous Models}
The performance characteristics in a heterogeneous scenario, i.e. with nonzero $\xi,\gamma$, are for the most part broadly similar to the homogeneous case, so we report here only on the observed differences. The overall classification performance both in terms of fraction correct and mutual information, is slightly improved for both independent and Ising decoders with heterogeneous neural ensembles. As an example, the accuracy as a function of ensemble size is compared for the basic model and the ``fully heterogeneous'' ($\gamma=\xi=1$) models in Fig.\ref{fig:6}, for both TAPwd and Independent decoders. The greater performance for nonzero $\gamma,\xi$ can be explained by the greater variability of cell properties, allowing more specific response patterns than in the homogeneous scenario. Such a scenario is presumably relevant to many real-world decoding problems, suggesting that decoder performance analysed with homogeneous test data may slightly under-represent real-world performance.

\begin{figure}[htb]
 \centering
 \includegraphics{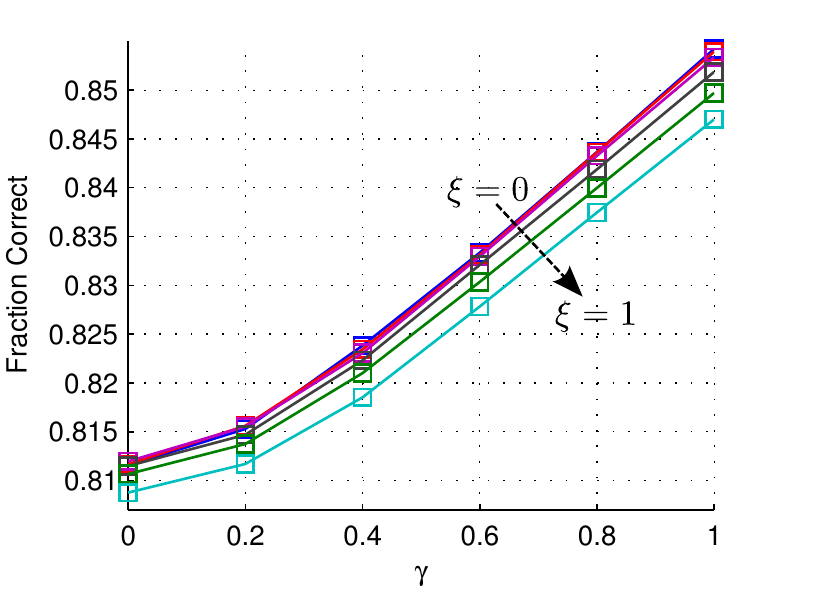}
  \caption{Decoder performance dependence on parameters $\xi,\gamma$. With increasing $\gamma$ the decoding performance is enhanced. The tuning width heterogeneity as specified by $\xi$ reduces the performance of the decoder (values from upper left to lower right curve $\xi = \{0,0.2,0.4,0.6,0.8,1\}$). Data for 70 cells, TAPwd algorithm used to train the decoder.}
 \label{fig:7}
\end{figure}

We also assessed the relative influence of the individual parameters $\gamma,\xi$ on the decoding behavior as shown in Figure \ref{fig:7}. Our analysis shows that, while the heterogeneity in the firing rates as specified by $\gamma$ has a positive effect on the decoding performance, an increased tuning width heterogeneity slightly decreases the performance of the decoder. However, the relative influence of the parameter $\xi$ is small.

\subsection{Dependence on level of correlation}

As the advantage of the Ising Decoder over the Independent Decoder stems from its ability to take advantage of information contained in pairwise correlations, we examined the dependence of this advantage on the average strength of correlation. Although we have set the average level of correlation to what has traditionally been thought to be a reasonable level for cells in the same neighbourhood \cite{Zohary1994}, there is an ongoing debate about the level and stimulus dependence of correlation relevant for cortical function \cite{Renart2010,Bair2001,Nase2003,Ecker2010}. This is of course critical for the performance of the Ising decoder. If there were no (noise) correlations present in the data, i.e. an independent decoder were the correct model for the data, there would be no benefit to using any decoder including correlation such as the Ising decoder. In fact, any decoder including correlations would most likely perform worse in practice than an independent decoder, as due to finite sampling effects one would most likely falsely estimate some small correlation in the data. Overfitting to these falsely learned correlations would then result in a performance decrease compared to an independent decoder, which by construction does not include any correlations, i.e. would implement the correct model.

\begin{figure}[tb]
 \centering
 \includegraphics{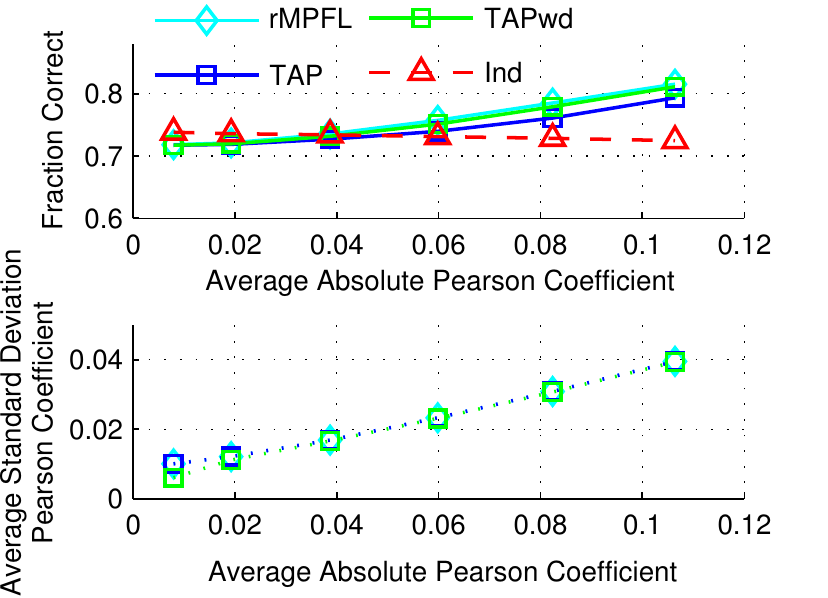}
 \caption{Ising decoder benefits from a correlation regime. Upper panel: Fraction correct as a function of the measured average absolute Pearson correlation coefficient for four decoding algorithms. Lower panel: Average Standard Deviation of the Pearson Correlation coefficient vs. average Pearson correlation coefficient. All averages taken over 20 simulations; 70 cells; basic model; correlations measured with 10000 samples.}
 \label{fig:8}
\end{figure}

This effect be seen in Figure \ref{fig:8}, where we have varied the mean absolute correlation strength as outlined in the methods section.
It can be seen that as the level of correlation increases for specified spike count, the independent decoder loses some discriminative capacity. The Ising decoders, however, take advantage of the higher level and spread of correlation values for discrimination between stimuli.
As described in Methods, at the lower end of this curve the underlying latent Gaussian has an identity correlation matrix, and thus the individual neurons are actually independent, although the average measured absolute correlation does not completely decrease to zero. This provides an explanation of why the Ising decoder fails to beat the independent decoder: by assuming the measured correlations are due to the true distribution, it overfits to these correlations, and thus performs worse compared to the (by construction correct) independent model. It should be noted, however, that the performance drop of the Ising decoder compared to the independent decoder is small even for a regime where cells are effectively independent.

To mimic the richer correlation structures potentially found in real neural recordings, we performed further testing. We simulated larger ensembles of neurons, while keeping the number of observed cells fixed, thus effectively creating "hidden" neurons. The visible neurons were simulated as described in methods, while for every unobserved cell, the preferred tuning direction was chosen randomly according to a uniform distribution. By using this scheme we could assure that the observed $C_\text{visible}$ cells always had the same characteristics except the correlation structure, which was effectively changed by the introduction of $C_\text{hidden}$ unobserved neurons. It should be noted that introducing hidden cells alters not only the second order statistics but also higher order correlations in the data, thus providing a much richer statistical structure in the data.

\begin{figure}[tb]
 \centering
 \includegraphics{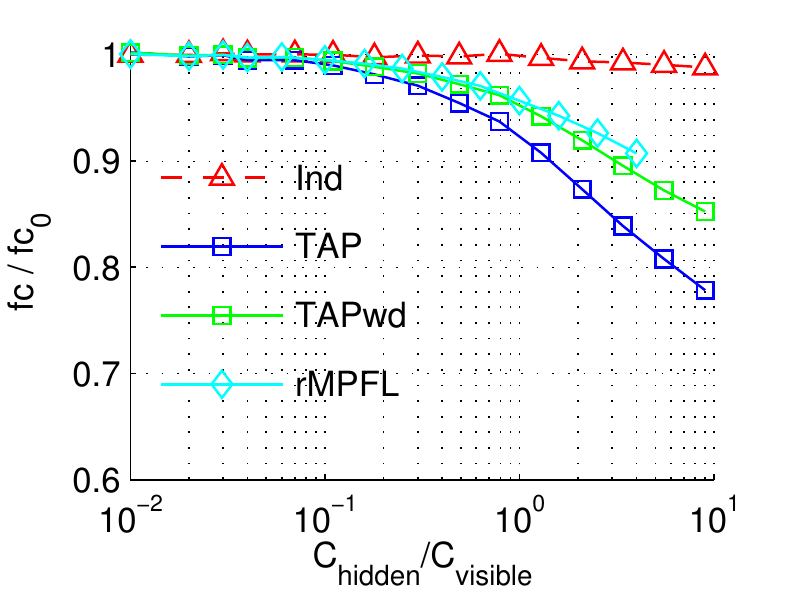}
 \caption{Decoder robustness to correlations due to hidden units. Fraction Correct $fc$ for 100 cells, normalized to the fraction correct $fc_0$ where 100 out of 100 cells are visible, is displayed as a function of the ratio between $C_\text{hidden}$ hidden cells and $C_\text{visible}$ observed cells. For simulations the basic model was used.}
 \label{fig:9}
\end{figure}

The different decoding (and training) methods vary in their robustness towards the introduction of such higher order correlation structure (Fig.~\ref{fig:9}). While the independent decoder is relatively unaffected by the introduction of additional unobserved cells, the Ising decoder model is more sensitive to such changes. However, significant differences between the different training strategies can be observed: in our case the rMPFL method showed the least performance drop (about 10\%) in a scenario where approximately 100 out of 500 cells were observed, while the fraction correct for the standard TAP approach drops by more than 15\% in this scenario. TAPwd is less affected than TAP, consistent with its inclusion of the leading terms for the next order expansion.

\subsection{Confusion Matrix Analysis}
\label{sec:confmat}

\begin{figure*}[tb]
 \centering
 \includegraphics{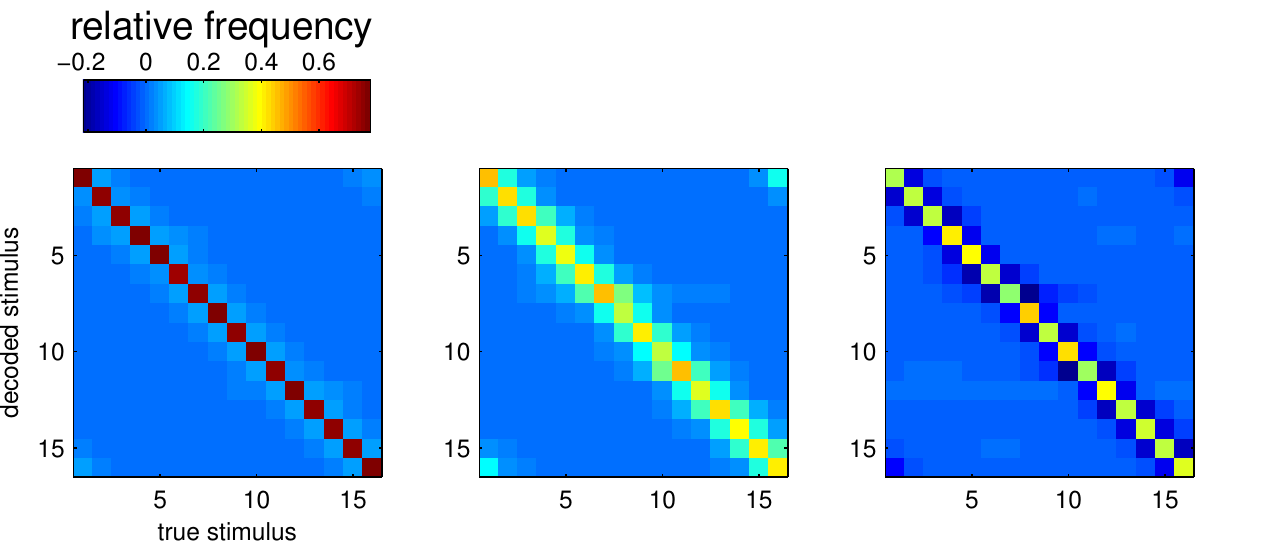}
 \caption{Decoder confusion matrices. The case illustrated is for 16 Stimuli, a 200-neuron ensemble, under the basic model ($\gamma=\xi=0$). Relative frequency as an approximation to the conditional probability with which each stimulus is decoded, for 10000 stimulus presentations. \textbf{a} Ising model decoder (TAPwd). \textbf{b} Independent decoder. \textbf{c} A difference plot between \textbf{a} and \textbf{b}. }
 \label{fig:10}
\end{figure*}

The confusion matrix provides complete information about the decoding error distribution. Confusion matrices for the Ising decoder and the Independent decoder respectively are shown in Fig.~\ref{fig:10}. The Ising decoder has higher diagonal terms, corresponding to better decoding accuracy. The overall appearance of the Ising decoder confusion matrix is fairly similar to the independent decoder. However, by comparing the two confusion matrices it can be seen that the Ising decoder mainly gains its performance benefit over the independent decoder by avoiding confusion of adjacent stimuli. This shows that as the difference in adjacent stimulus directions becomes less with an increasing number of stimuli, the Ising model decoder can utilize the correlation patterns to enhance the decoding accuracy, i.e. to distinguish between adjacent stimulus directions more precisely. This effect of course depends on the correlation model used - here the correlation between each pair of neurons was resampled for each stimulus in the simulation. If (noise) correlations were not at all stimulus-dependent, or if the model was quite different, then the Ising decoder may not be able to take advantage of this potential performance advantage.

\subsection{Comparison with linear decoding}

While most work in the Brain-Machine Interface literature has focused on continuous decoders, there has been some work on the discrete decoding problem, although to date with a focus on the analysis of either continuous data, such as electroencephalographic (EEG) data \cite{Wolpaw2002}, or on longer time windows of multi-electrode array data, in which spike counts are far from binary \cite{Santhanam2009}. As the discrete decoding problem can be viewed as a classification problem (in the same sense as the continuous decoding problem can be seen as a regression problem), it is of interest to compare the performance of our approach with traditional classification approaches such as the Optimal Linear Classifer (OLC).

Following \citeasnoun{Bishop2007}, each stimulus class can be described by its own linear model, so that
\begin{equation}
 y_k = \mathbf{w}_s^T\vec\sigma + w_{k_0},
\end{equation}
where $s=1,\ldots,S$.
Using a 1-of-S binary coding scheme (i.e. we denote stimulus class $s_i$ by a ``target'' column vector $\mathbf{t}$ with all zeros except the i-th entry, which is one) the weights $\mathbf{w}_s$ can be trained such as to minimize a sum of squares error function for the target stimulus vector.

\begin{figure}[bt]
 \centering
 \includegraphics{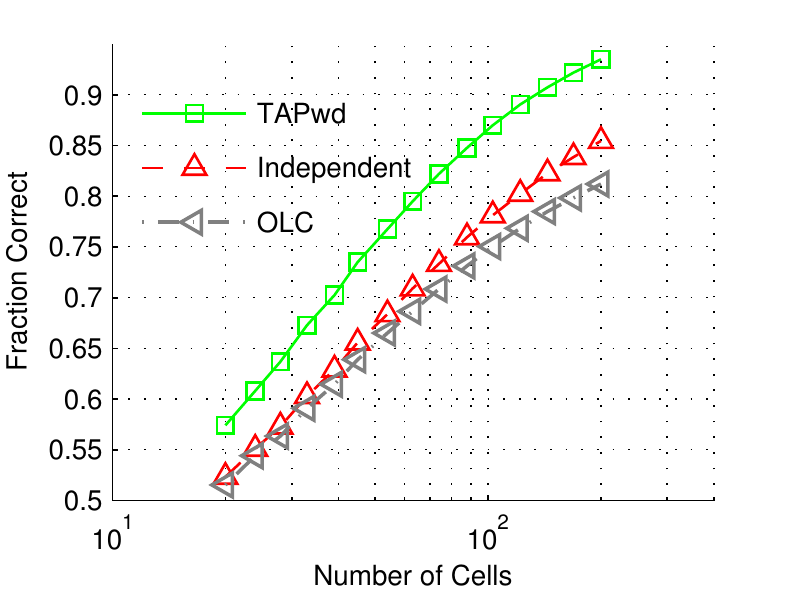}
 \caption{Comparison of the TAPwd Ising Decoder and Independent Decoder with the Optimal Linear Classifier (OLC).}
 \label{fig:11}
\end{figure}

This is done in Fig.~\ref{fig:11}. It can be seen that, under the conditions we test here (10000 trials, simulated data as described previously for the basic model), the OLC underperforms both the Ising (as exemplified by the TAPwd approach) and Independent classifiers. The former is not unexpected, but it may seem initially counter-intuitive that  the OLC does not yield identical performance to the Independent decoder, as the latter is in effect performing a linear classification.

However, the following differences must be noted for these two algorithms.  As their implementation details differ, they may have markedly different sensitivity to limited sampling - the independent decoder, as we have constructed it here (product of marginals) has remarkable sampling efficiency. Most importantly however, the OLC decoder assumes by construction a Gaussian error in the stimulus class target vectors, i.e. it corresponds to a maximum likelihood decoding when assuming that the target vectors follow a Gaussian conditional distribution \cite{Bishop2007}. This assumption is clearly not valid in the case of binary target vectors. Therefore the failure of the OLC decoder should not be surprising.

\section{Discussion}

We have demonstrated, for the first time, the use of the Ising model to decode discrete stimulus states from simulated large-scale neural population activity. To do this, we have had to overcome several technical obstacles, namely the poor scaling properties of previously used algorithms for learning model parameters, and similarly the poor scaling behavior of methods for estimating the partition function, which although not necessary for some applications of the Ising model in neuroscience, is required for decoding. The Ising model has one particular advantage over a simpler independent decoding algorithm: it can take advantage of stimulus dependence in the correlation structure of neuronal responses, where it exists. With the aid of a statistical simulation of neuronal ensemble spiking responses in the mouse visual cortex, we have demonstrated that correlational information can be taken advantage of for decoding the activity of neuronal ensembles of size in the hundreds by several algorithms, including mean field methods from statistical physics and the rMPFL algorithm.

Ising models have gained much attraction recently in neuroscience to describe the spike train statistics of neural ensembles \cite{Schneidman2006,Shlens2006,Shlens2009}. However, these findings have largely been made only in relatively small neural ensembles (typically a few tens of cells), from which an extrapolation to larger ensemble sizes might not be wise \cite{Roudi2009a,Roudi2009}. The principal reason for this limit has been the poor scaling of the computational load of fitting the Ising model parameters, when algorithms such as Boltzmann learning are used.
Moreover, new findings suggests that pairwise correlations (and thus Ising models) might not be sufficient to predict spike patterns of small scale local clusters of neurons ($ < 300\mu m$ apart), which have been observed to provide evidence of higher order interactions \cite{Ohiorhenuan2010}.
While the formalism for higher order models may be similar, scaling properties are guaranteed to be even worse. There is thus the pressing need to develop better algorithms for learning the parameters of Ising and Ising-like models.
It should be noted that while Ising models might not necessarily provide an exact description of neural spike train statistics, for decoding purposes we only require that we can approximate their distribution well enough to achieve good decoding performance.

The development of discrete neural population decoding algorithms has two motivations. The first motivation is the desire to develop brain-computer communication devices for cognitively intact patients with severe motor disabilities \cite{Mak2009}. In this type of application, an algorithm such as those we describe could be used together with multi-electrode brain recordings to allow the user to select one of a number of options (for instance a letter from a virtual keyboard), or even in the longer term to communicate sequences of symbols from an optimized code directly into a computer system or communications protocol. Given the short timescale to which the Ising decoder can be applied (we have fixed this at 20 ms here), and sufficiently large recorded ensembles to saturate decoder performance, very high bit rates could potentially be achieved.

The second motivation is more scientific: to use such decoding algorithms to probe the organization and mechanisms of information processing in neural systems. It should be immediately be apparent that the Ising decoder and related models can be used to ask questions about the neural representation of sensory stimuli, motor states or other behavioral correlates, by comparing decoding performance under different sets of assumptions (for instance, by changing the constraints in an Ising model to exclude correlations, include correlations within 50 $\mu$m, etc). This (commonly referred to as the ``encoding problem") is essentially the same use to which Shannon information theory has been applied in neuroscience (see e.g.  \citeasnoun{Schultz2009} for a recent example), with simply a different summary measure.
Use of decoding performance may be an intuitively convenient way to ask such questions, but it is still asking exactly the same question. However, there are other uses to which such algorithms can be applied. For instance, combining sensory and learning/memory experimental paradigms, once a decoder has been trained, it could be used subsequently to read out activity patterns in different brain states such as sleep, or following some period of time - for instance, to "read out memories" by decoding the patterns of activity that represent them. The decoding approach may thus have much to offer the study of information processing in neural circuits.

Our results show that decoding performance is critically dependent on the sample size used for training the decoder, as relatively precise characterization of pairwise correlations is needed to fit a model that matches the statistical structure of as-of-yet unobserved data well. In the ``encoding problem'', such finite sampling constraints result in a biased estimate of the entropy of the system.
For the decoding problem considered here, finite sampling leads to overfitting of the model to the observed training data, so that it does not generalize well to the unobserved data, and accordingly fails to predict stimulus classes correctly during test trials. Such finite sampling constraints mean that below a particular sampling size - which we found to be 400 trials for 70 neurons in one particular example we studied - there is no point in using a model which attempts to fit pairwise (or above) correlations, one may as well just use an independent model.
This has implications for experimental design. However, it should be noted that the real brain has no such limitation - in effect, many thousands or millions of trials are available over development, and so a biological system should certainly be capable of learning the correct correlations from the data \cite{Bi2001} and thus may well be able to operate in a regime where decoding benefits substantially from known correlation structure.

We have shown that incorporating correlations in the decoding process might be especially relevant for `hard' decoding problems, i.e. multi-class discrimination problems in which stimuli are not easily distinguishable by just observing individual neuronal firing rates. In this scenario including correlations could be a means to enhance the precision of the decoding process by increasing the discriminability between adjacent or similar stimuli. Including correlations could make the pattern distribution more flat, or uniform, with low firing rates, leading to greater energy efficiency of information coding \cite{Baddeley1997}.

We must insert a note of caution concerning the presence of higher order correlations in data to be fitted. Such higher order correlations might arise simply from the presence of a large number of "hidden" neurons whose activity has not been recorded, but which have a substantial influence on the activity of those neurons recorded. This is likely to occur frequently in real neurophysiological situations. The performance of pairwise correlation decoder models, such as the Ising model, is necessarily affected detrimentally by such effects, as shown in Fig. 8. Interestingly, an independent decoder is less affected (although it may also be capturing less information anyway). Obviously, it is possible to make use of higher order models to alleviate this problem, but a penalty then has to be paid in sampling terms.

We note that the primary use of the Ising model in neuroscience so far has been to model the empirical statistics of neural spike train ensembles \cite{Schneidman2006,Shlens2006}. There is of course no requirement that a good decoder is also a good model of neural spike train statistics - what matters is its performance on the test dataset. Nevertheless, knowing how well the model captures spike train pattern structure may help to build better decoders, and of course may be of particular value when those decoders are used to study neural information processing (as opposed to being used for a practical purpose such as brain-machine interface development). Unfortunately, a direct comparison of empirical to model spike pattern probabilities is not experimentally feasible for large ensemble sizes - while this is relatively simple for 15 cells, it is far from viable for 500 cells. Further work is needed to determine how best to evaluate the performance of decoders at predicting empirical spike patterns, when decoding very large neural ensembles.

One caveat to the advantage provided by correlations of using Ising over Independent decoders is that it depends entirely upon the extent to which correlations are found to depend upon the stimulus variables of interest. While a previous study at longer timescales has found correlations to improve neural decoding \cite{Chen2006}, the jury is still out on the prevalence of stimulus dependence of pairwise and higher order correlations in the mammalian cortex. Stimulus-dependent correlations have been found in mouse \cite{Nase2003}, cat \cite{DasGilbert1999,Samonds2005} and monkey \cite{Kohn2005,Kohn2009}, where they have been shown to contribute to information encoding \cite{Montani2007}, but most recordings to date have sampled relatively sparsely from the local cortical circuit, due to limitations in multi-electrode array hardware.
It is possible that if one were to be able to record from a greater proportion of neurons in the local circuit, then stronger stimulus-dependent correlations might be observable.

In an experimental setting the performance could also be enhanced by choosing an optimal timebin-width, a question that we have not addressed in this paper.
However care is needed for choosing the right bin-width. As shown by Roudi et al. \cite{Roudi2009,Roudi2009a}, a small bin-width is likely to yield a good fit, however choosing a too small bin-width invalidates the underlying assumption of uncorrelated time-bins. Choosing a too large time-bin makes it however harder to find a good fit for the model and additionally may violate the assumptions of binary responses.

A number of avenues present themselves for future development of decoding algorithms. Firstly, algorithms for reducing model dimensionality without losing discriminatory power, may prove advantageous. These may include graph and hypergraph theoretic techniques \cite{Aghagolzadeh2010} for pruning out uninformative dimensions (edges and nodes), and factor analysis methods for modeling conditional dependencies \cite{Santhanam2009}. Such an approach may be particularly advantageous when experimental trials are limited, as the dimensionality of the parameter set is the main reason for Ising model performance not exceeding the Independent model for limited trials. One difficulty with the use of graph pruning approaches is that the usual pairwise correlation matrix of neural recordings, unlike the graph in many network analysis problems, tends not to be sparse. It is of course a functional, as opposed to synaptic, connectivity matrix, and one reason for this lack of sparseness is its symmetric nature. It has recently been proposed that the symmetry property of the $J_{ij}$ matrix can be relaxed in the context of the (non-equilibrium) Kinetic Ising model, which also provides a convenient way to take into account space-time dependencies, or causal relationships \cite{Hertz2010,Roudi2011}. Use of the Kinetic Ising model framework for decoding would appear to be an interesting future direction to pursue.

New experimental technologies are yielding increasingly high dimensional multivariate neurophysiological datasets, usually without concomitant increases in the duration of data that can be collected. However, there is some reason for optimism that we will be able to develop new data analysis methods capable of taking advantage of this data. Maximum entropy approaches to the fitting of structured parametric models such as the Ising model and its extensions would appear to be one approach likely to yield progress.

\subsubsection*{Acknowledgements}
We thank Phil Bream, H\'el\`ene Seiler and Yang Zhang for their contributions to earlier work leading up to that reported here, and Aman Saleem for useful discussions and comments on this manuscript. We also thank Yasser Roudi for useful comments on the TAP equations, and Jascha Sohl-Dickstein, Peter Battaglino, and Michael R DeWeese for useful MATLAB code and discussion of the MPFL technique. This work was supported by EPSRC grant EP/E002331/1 to SRS.

\end{document}